\documentclass[
preprint,11pt
]{elsarticle}

\usepackage{amssymb}
\usepackage{amsmath}
\usepackage{multirow}
\usepackage{rotating}
\usepackage{graphicx}
\usepackage{dcolumn}
\usepackage{bm}
\usepackage{epstopdf}

\journal{Physica A}
\begin{document}

\begin{frontmatter}

\title{Architecture of the Florida Power Grid as a Complex Network}

\author[]{Yan Xu\corref{cor}}
\ead{yxu2@fsu.edu} 
\author{Aleks Jacob Gurfinkel}
\ead{agurfinkel@fsu.edu}
\author{Per Arne Rikvold}
\ead{prikvold@fsu.edu}

\cortext[cor]{Corresponding Author. Tel.: +1 8506446176}

\address{Department of Physics, Florida State University, Tallahassee, FL 32306-4305, USA}

\date{\today}

\begin{abstract}
We study the Florida high-voltage power grid as a technological network embedded in space. Measurements of geographical lengths of transmission lines, the mixing of generators and loads, the weighted clustering coefficient, as well as the organization of edge conductance weights show a complex architecture quite different from random-graph models usually considered. In particular, we introduce a parametrized mixing matrix to characterize the mixing pattern of generators and loads in the Florida Grid, which is intermediate between the random mixing case and the semi-bipartite case where generator-generator transmission lines are forbidden. Our observations motivate an investigation of optimization (design) principles leading to the structural organization of power grids. We thus propose two network optimization models for the Florida Grid as a case study. Our results show that the Florida Grid is optimized not only by reducing the construction cost (measured by the total length of power lines), but also through reducing the total pairwise edge resistance in the grid, which increases the robustness of power transmission between generators and loads against random line failures. We then embed our models in spatial areas of different aspect ratios and study how this geometric factor affects the network structure, as well as the box-counting fractal dimension of the grids generated by our models. 
\end{abstract}


\end{frontmatter}

\section{Introduction}

A power grid is defined as a network of high-voltage (100-1000 kV) transmission lines that provide long-distance transport of electric power within and between countries; low voltage lines that provide local power delivery are normally excluded. It is a spatially embedded technological network responsible for power generation and transmission. The vertices in a power-grid network correspond to generating stations (generators) and switching or transmission substations (loads), while the edges correspond to the high-voltage lines between vertices. Over recent years, complex network analysis has proven useful to the study of networked systems in technology, sociology, biology, and information science \cite{siam, Newman}. Given that today's electrical grids are the largest engineered systems ever built \cite{gripgrid}, as well as the emergence of transdisciplinary electric power-grid science \cite{GridSci}, we study the Floridian (high-voltage) power transmission grid as a complex network \cite{Sun2005, GridConf, gridnetwork} by characterizing its topology and investigating its structural architecture. The purpose of our investigation is to explore organizational principles of power grids, using the Floridian high-voltage grid as a case study. 

The organization of this paper is as follows. In Section II, we view the Florida Grid as a spatial network, focusing on the structural organization of geographic lengths and edge conductance weights, as well as the mixing patterns of generators and loads. In Section III, we propose network optimization models that exhibit desirable properties, and show that they coincide with the real Florida Grid on several key network measures. In Section IV, we compare the box-counting fractal dimensions of the power-grid networks generated by our models. Our conclusions and some suggestions for future research are given in Section V.

\section{The Floridian power grid as a complex weighted network}
\begin{figure}
  \begin{center}    
  \hspace*{-1cm}\includegraphics[height=10cm]{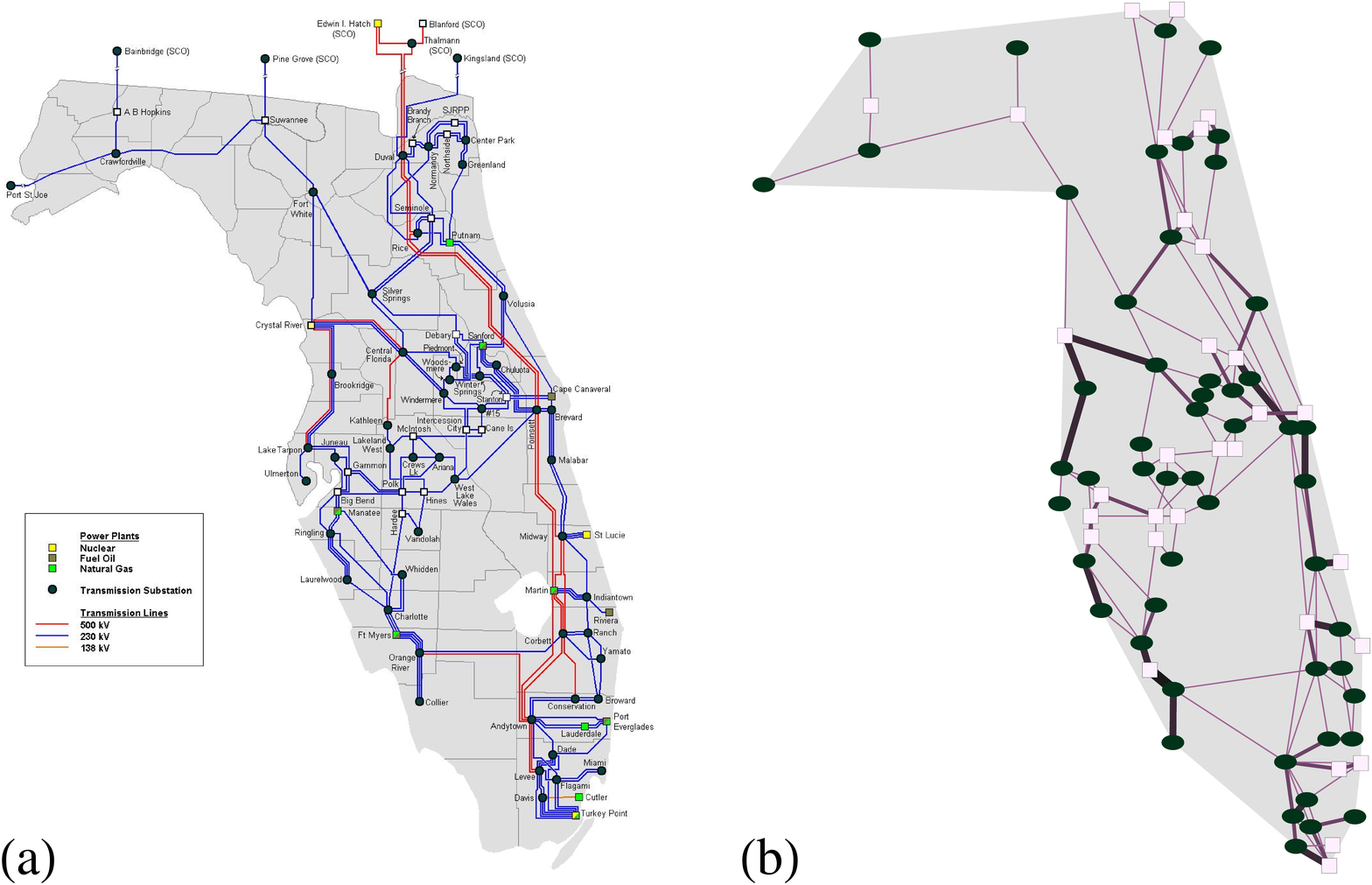} 

\end{center}
\caption{The Florida Grid and its network representation. (a) Map of the Floridian high-voltage grid \cite{map}. Despite being referred to as the ``Florida Grid" (FLG), this map does not cover power stations in the far northwest corner of Florida state, while including 2 power plants and 4 transmission substations in southern Georgia. (b) FLG in our network representation, in which power lines are represented as straight lines between vertices. The distances are normalized such that the number of vertices per unit area is unity, and the total area is illustrated as the shaded region. Generators are represented as squares and loads as ovals. Thicker, darker links represent multiple power lines between vertices.}
\label{FloridaGrid}
\end{figure}

The Floridian high-voltage power grid (hereafter referred to as ``FLG", see map in Fig. \ref{FloridaGrid}), is a relatively small network consisting of $N=84$ vertices ($N_g=31$ generators and $N_l=53$ loads). The vertices are connected by $M=200$ power transmission lines. Since there are often multiple lines between stations in FLG, for clarity, we use the following nomenclature. ``Edge'': any connection between vertices $i$ and $j$. ``Transmission line": any single physical power transmission line or its network representation. Parallel transmission lines can be seen in Fig. \ref{FloridaGrid}(a). ``Multiple edge": any edge corresponding to more than one parallel transmission line, as represented by the thicker links in Fig. \ref{FloridaGrid}(b). 


Here we define the \emph{degree} to be the number of transmission lines connecting to a vertex. The average degree of the grid is $\langle k \rangle \equiv 2M/N \approx 4.76$, while the average degrees of generators and loads are  $\langle k_g \rangle \approx 4.48$ and $\langle k_l \rangle \approx 4.93$, respectively. Note that $\langle k_g \rangle < \langle k \rangle < \langle k_l \rangle$, i.e., there are on average more transmission lines connecting to loads than to generators.

\subsection{Geographic length distribution of power lines}

In order to characterize the spatial properties of the grid in a straightforward and consistent way, we represent FLG using distance units that are normalized such that the number of vertices per unit area is unity \cite{Rikvold}. This is accomplished by confining the $N$ vertices to a spatial region of area $N$. See Fig. \ref{FloridaGrid} for illustration and also the square area case described in the beginning of Section III as an example.

Here we measure (in these units) the average length of all lines $\langle\ell\rangle$, the average length of lines that connect generators to generators $\langle\ell_{gg}\rangle$, the average length of lines connecting generators with loads $\langle\ell_{gl}\rangle$, and the average length of lines joining two loads together $\langle\ell_{ll}\rangle$. The results are $\langle\ell\rangle \approx 1.09$, and $\langle\ell_{gg}\rangle \approx 0.71 < \langle\ell_{gl}\rangle \approx 1.02 < \langle\ell_{ll}\rangle \approx 1.27$, indicating that power transmission lines around loads are on average longer than those around generators. In particular, we see that the power lines between generators and generators are typically much shorter than the overall average value.

\subsection{Mixing patterns of generators and loads}

Generators and loads play different roles in power grids, so it is of interest to investigate the pattern of their connections to each other. Assortative (or disassortative) mixing is the preference of a network's vertices to attach to others that are similar (or dissimilar) in some way. Such mixing can be characterized by the elements of a \emph{mixing matrix} $[e_{ij}]$ \cite{Mixing}, defined to be the fraction of lines in a network that connect a vertex of type $j$ (tail) to one of type $i$ (head). The mixing matrix satisfies the sum rule $\sum_{ij} {e_{ij}}=1$. Following the definitions in \cite{Mixing}, $a_i \equiv \sum_{j} {e_{ij}}$ quantifies the fraction of lines whose \emph{heads} are attached to vertices of type $i$. Similarly, $b_j \equiv \sum_{i} {e_{ij}}$ quantifies the fraction of lines whose \emph{tails} are attached to vertices of type $j$. For undirected networks (in which there is no distinction between head and tail for lines), e.g., power grids, the mixing matrix is always symmetric, $e_{ij}=e_{ji}$, and thus $a_i=b_i$. In power grids, there are only two types of vertices, viz. generators ($g$) and loads ($l$), thus leading to a $2 \times 2$ mixing matrix:
\noindent
\begin{equation}
[e_{ij}] = \begin{pmatrix}
       e_{gg} & e_{gl}           \\[0.3em]
	e_{gl}  & e_{ll}          
     \end{pmatrix} ,
\end{equation}   %
where $e_{gg}$ denotes the fraction of transmission lines that connect generators to generators, with similar meanings for  $e_{gl}$ and $e_{ll}$. The mixing matrix of FLG is
\noindent
\begin{equation}
[e_{ij}]_{\rm{FLG}} = \begin{pmatrix}
       0.085    &   0.2625           \\[0.3em]
       0.2625  &   0.39    
     \end{pmatrix}.
\end{equation}

An \emph{assortativity coefficient} $r$ is defined to quantify the level of assortative mixing in networks \cite{Mixing}. For power grids (undirected networks), 
\noindent
\begin{equation}
r=\frac{\sum_{i}{e_{ii}}-\sum_{i}{{a_i}^2}}{1-\sum_{i}{a_i}^2}=\frac{e_{gg}+e_{ll}-{(e_{gg}+e_{gl})}^2-{(e_{gl}+e_{ll})}^2}{1-{(e_{gg}+e_{gl})}^2-{(e_{gl}+e_{ll})}^2} .
\end{equation}
This yields $r_{\rm{FLG}} \approx -0.158$, the sign indicating that generators and loads in FLG are disassortatively mixed. 

The densities of generators and loads in a power-grid network are $\rho_g=N_g/N$ and $\rho_l=N_l/N$, respectively. For FLG, $\rho_g=31/84 \approx 0.369$ and $\rho_l=1-\rho_g \approx 0.631$. For comparison, we also estimate the mixing matrix that would result if vertices were randomly mixed with the same densities as in FLG:
\noindent
\begin{equation}
[e_{ij}]_{\rm{rand}} = \begin{pmatrix}
       \rho_g^2    &   \rho_g \rho_l           \\[0.3em]
       \rho_g \rho_l   &     \rho_l^2
     \end{pmatrix}
\approx \begin{pmatrix}
       0.136    &   0.233          \\[0.3em]
       0.233    &   0.398
     \end{pmatrix}.
\end{equation}
Comparing the ``real'' (FLG) and ``random'' (rand) mixing matrices, we immediately see that $(e_{gg})_{\rm{FLG}}=0.085<(e_{gg})_{\rm{rand}} \approx 0.136$, $(e_{gl})_{\rm{FLG}}=0.2625>(e_{gl})_{\rm{rand}} \approx 0.233$, and $(e_{ll})_{\rm{FLG}}=0.39$ is very close to $(e_{ll})_{\rm{rand}} \approx 0.398$. These results indicate that in FLG, generator-generator ($g-g$) connections are disfavored, while generator-load ($g-l$) connections are favored. Load-load ($l-l$) connections are almost unchanged with respect to random mixing. This observation can be understood by considering random failures of generators and loads as follows.


In power grids, generators produce energy, while loads play the roles of both consumer and switching station, i.e., a load A may transfer energy from an adjacent generator G to another load B, which is not directly connected to any generators. The failure of a switching load A can thus affect the power supply of another load B. Hence, from a practical viewpoint, in power grids loads are preferred to have direct connections with generators. This design preference is reflected by the fraction of $g-l$ lines being higher than that in the random case: $(e_{gl})_{\rm{real}}>(e_{gl})_{\rm{rand}}$. At the same time, a generator is self-sufficient in producing energy and thus does not have to be joined via direct transmission lines to another generator, so that $(e_{gg})_{\rm{real}}<(e_{gg})_{\rm{rand}}$. 

This discrepancy of mixing patterns between FLG and its random mixing counterpart can be visualized in an intuitive way. Starting from the random mixing matrix, due to the technological considerations discussed above, a significant fraction of the diagonal element $e_{gg}$ flows to the off-diagonal elements $e_{gl}$ and $e_{lg}$ (here $e_{lg}=e_{gl}$), while the other diagonal element $e_{ll}$ is almost unaffected. Note that during this matrix-element ``migration'' process, the sum rule $e_{gg}+2e_{gl}+e_{ll}=1$ is preserved. By decreasing $e_{gg}$ and increasing $e_{gl}$ in this manner, we can approximate the mixing pattern of generators and loads in FLG. Here we introduce a parameter $\alpha$ ($0 \leq \alpha \leq 1$) quantifying the fraction of the diagonal element, $e_{gg}$, migrated to the off-diagonal ones. Hence, a mixing matrix as a function of $\alpha$ and $\rho_g$ is obtained (recall that $\rho_g+\rho_l=1$ by definition):
\noindent
\begin{eqnarray}
[e_{ij}(\alpha, \rho_g)]&= &  \begin{pmatrix}
       (1-\alpha)\rho_g^2    &   \rho_g \rho_l + \frac{\alpha}{2} \rho_g^2     \\[0.3em]
       \rho_g \rho_l + \frac{\alpha}{2} \rho_g^2   &     \rho_l^2
     \end{pmatrix}  \nonumber \\[0.3em]
&=&\begin{pmatrix}
       (1-\alpha)\rho_g^2    &   \rho_g - (1-\frac{\alpha}{2}) \rho_g^2         \\[0.3em]
       \rho_g - (1-\frac{\alpha}{2}) \rho_g^2    &   (1-\rho_g)^2
     \end{pmatrix},
\end{eqnarray}
with assortativity coefficient
\noindent
\begin{equation}
r(\alpha,\rho_g)=-\frac{\alpha \rho_g (4-4\rho_g+\alpha {\rho_g}^2)}{(2-2\rho_g+\alpha {\rho_g}^2)(2-\alpha \rho_g)}.
\end{equation}
For $\alpha=0$, which corresponds to the random mixing case, $e_{ij}(0,\rho_g)= (e_{ij})_{\rm{rand}}$, and $r(0,\rho_g)=r_{\rm{rand}}=0$.
For  $\alpha=1$, $e_{gg}=0$, viz. generators are not allowed to have direct connections with each other, and we call this the ``semi-bipartite'' (semi-bip) case \footnote{ 
For FLG with $\rho_g=31/84$, the semi-bipartite mixing matrix becomes
\noindent
\begin{equation*}
[e_{ij}]_{\rm{semi-bip}} \approx \begin{pmatrix}
      0    &  0.301         \\[0.3em]
      0.301     &   0.398
     \end{pmatrix},
\end{equation*}
with assortativity coefficient $r_{\rm{semi-bip}} \approx -0.431$. 
}

\noindent
\begin{eqnarray}
[e_{ij}]_{\rm{semi-bip}}=[e_{ij}(\alpha=1,\rho_g)]&=& \begin{pmatrix}
      0    &   \rho_g - \frac{1}{2} \rho_g^2        \\[0.3em]
      \rho_g - \frac{1}{2} \rho_g^2   &     (1-\rho_g)^2
     \end{pmatrix}.
\end{eqnarray}

\newpage
The mixing matrix of FLG is intermediate between the random mixing case ($\alpha=0$) and the semi-bipartite case ($\alpha=1$). Letting  $e_{gg}(\alpha, \rho_g=31/84)=(e_{gg})_{\rm{FLG}}=0.085$ and solving  for $\alpha$, we obtain $\alpha_{\rm{FLG}} \approx 0.375$. This $\alpha$ value simply means that compared to the random mixing of generators and loads, about $37.5\%$ of $g-g$ connections are suppressed and replaced by $g-l$ connections in FLG. Interestingly, this value of $\alpha_{\rm{FLG}}$ is very close to the density of generators in FLG, viz. $\alpha_{\rm{FLG}} \approx 0.375$ is very close to $\rho_g \approx 0.369$. Analysis of more real-world power grids would be needed in order to specify whether this is a coincidence that holds for FLG only, or due to some possible universal pattern for power grids.

\subsection{Conductance edge weight and transmission line redundancy}

In power grids, there are often multiple transmission lines between a given pair of stations. This increases redundancy (e.g., in case one of the lines fails, there is still some direct transmission capacity available), and reduces the effective resistance between such vertices. Assuming that the electrical conductance between two vertices is proportional to the multiplicity of lines and inversely proportional to the corresponding geographical distance, we define an ``electrical conductance weight'' associated with an edge between vertices $i$ and $j$ to be \cite{Hamad}
\noindent
\begin{equation}
w_{ij} \equiv \frac{\textrm{number of direct transmission lines connecting vertices \em i \em and \em j \em}}{\textrm{geographical distance between vertices \em i \em and \em j \em}}
\label{weight} .
\end{equation}\vspace{0em}
 
\noindent Hence the power grid can be further characterized by an $N \times N$ symmetric \emph{conductance matrix}, $\bf{W}=$ $[{w_{ij}}]$ \cite{Hamad}. Clearly, $\bf{W}$ is a \emph{weighted} version of the \emph{adjacency matrix} $\bf{A}=$ $[A_{ij}]$ ($A_{ij}=1$ if vertices $i$ and $j$ are connected by an edge, and $A_{ij}=0$ otherwise \cite{Newman}). We define the total pairwise \emph{resistance}, viz. the sum of edge resistances (inverse edge conductances),
\noindent
\begin{equation}
R \equiv \sum_{ij} {A_{ij} w_{ij}}^{-1}
\label{R},
\end{equation}
where the adjacency matrix element $A_{ij}$ ensures that only pairs of adjacent vertices $ij$ contribute in the summation. For FLG, the total pairwise resistance is measured to be $R \approx 139.78$ in our units. 

The edge weight $w_{ij}$ has a twofold meaning. One is inverse edge resistance, another is robustness in the following sense. The probability of random failure on an edge is proportional to the geographic length of that edge, so shorter edges are more robust against random line failures than longer ones. Also, multiple edges, consisting of more than one parallel transmission lines, increase redundancy and thus reduce the risk that random line failure will completely sever a connection, so higher multiplicity of edges leads to more reliable power transmission between stations. From this point of view, it is natural to interpret $w_{ij}$ as an approximate measure of the robustness of the direct connection between vertices $i$ and $j$ against random failure of transmission lines. 

FLG has $137$ pairs of adjacent vertices (viz., 137 edges), thus there are $M_{\rm{ex}}=200-137=63$ extra transmission lines, which are placed as parallels. In other words, the average edge multiplicity is $200/137 \approx 1.46$, i.e., there are on average about $1.46$ transmission lines per edge. Of the 137 edges in FLG, 48 (about 35\%) are multiple edges, consisting of parallel transmission lines. Such abundance of multiple edges greatly reduces the total edge resistance in FLG. It is interesting to explore how resistances are organized among edges connecting vertices of both the same and different types. The average resistances between $g-g$, $g-l$ and $l-l$ adjacent vertex pairs are $\langle R_{gg} \rangle \approx 0.57$, $\langle R_{gl} \rangle \approx 0.96$ and $\langle R_{ll} \rangle \approx 1.20$, respectively. This organization pattern of resistance, $\langle R_{gg} \rangle < \langle R_{gl} \rangle < \langle R_{ll} \rangle$, is consistent with the arrangement of geographic lengths in FLG: $\langle \ell_{gg} \rangle < \langle \ell_{gl} \rangle < \langle \ell_{ll} \rangle$. We note in passing that the average multiplicities for $g-g$, $g-l$ and $l-l$ edges are quite similar, viz. about 1.42, 1.48 and 1.44, respectively.

\subsection{Weighted clustering and the organization of transmission lines}

The clustering coefficient $C$ measures the cliquishness of a typical neighborhood in networks. Clustering is an important measure in power grids, and has been of interest to researchers since the beginning of the study of complex networks \cite{WS}. Information on local connectedness is provided by the local clustering coefficient $c_i$, defined for any vertex $i$ as the fraction of connected neighbors of $i$ \cite{Newman},
\noindent
\begin{equation}
c_{i} \equiv \frac{\textrm{number of pairs of neighbors of \em i \em that are connected}}{\textrm{number of pairs of neighbors of \em i \em}}
\label{localclustering} .
\end{equation}
The average clustering coefficient $C=N^{-1} \sum_i {c_i}$ thus expresses the statistical level of cohesiveness, measuring the global density of interconnected vertex triples (triangles) in networks. Since FLG is a weighted network characterized by the edge conductance matrix $\bf{W}$, a natural definition of the local weighted clustering coefficient for vertex $i$ is \cite{Barrat2004}
\noindent
\begin{equation}
c_i^w \equiv \frac{1}{s_i (\tilde{k}_i-1)} \sum_{j,h} \frac{w_{ij}+w_{ih}}{2} A_{ij} A_{ih} A_{jh} ,
\label{weightedclustering}
\end{equation}
where $A_{ij}$ is an element of the (unweighted) adjacency matrix, $\tilde{k}_i$ is the number of edges connecting to vertex $i$, and $s_i \equiv  \sum_{j} A_{ij} w_{ij}$ is the vertex \emph{strength} of $i$ \cite{Barrat2004}. The average weighted clustering coefficient is then $C^w=N^{-1} \sum_i {c_i^w}$. 

The unweighted (substituting $w_{ij}$ with $A_{ij}$ in Eq. (\ref{weightedclustering})) and weighted clustering coefficients for the Florida power-grid network are $C_{\rm{FLG}} \approx 0.216$ and $C^w_{\rm{FLG}} \approx 0.213$, respectively. For comparison, the clustering coefficient for the Erd\H os-R\' eyni random graph \cite{Newman} (with the same $N$ and $M$ as in FLG) is $C_{\rm{ER}}=\langle k \rangle /(N-1)=2M/N(N-1) \approx 0.057$. Thus FLG is highly clustered, and such a high density of interconnected vertex triples makes FLG more robust against random failures of both vertices and edges. 

It is worth noting that the multiplicity of edges and the clustering behavior are interrelated. The total number of transmission lines in FLG is $M=200$, and they are placed among $N=84$ vertices. The minimum number of lines needed to make the whole network connected (as a spanning tree \cite{Newman}) is $N-1=83$ lines, so there are $M-(N-1)=200-83=117$ ``redundant'' transmission lines. The question becomes how to distribute a fixed number of redundant lines in a spatial network. One possibility is to connect more pairs of vertices, thus increasing the density of closed loops in the network, so that the clustering coefficient (which measures the density of loops of length 3, viz., triangles) also increases.  Alternatively, those redundant lines can be placed as parallels forming multiple edges between already connected pairs of vertices. Hence, there is a competition between clustering and edge multiplicity. Higher clustering means relatively fewer multiple edges, and vice versa. In the following sections, we will consider this competition in further detail.

\section{Spatial network optimization models}

In order to study the structural organization and architecture of FLG, we construct random-graph models for comparison, from which we see that certain optimization principles lead to desired network properties. We start from the null model $G(N,M)$, in which the number of vertices $N$ and transmission lines $M$ are fixed to be the same as those in FLG  (see Fig. \ref{cooling}(a)). First, $N=84$ vertices (including $N_g=31$ generators and $N_l=53$ loads) are positioned randomly in a square of side $\sqrt{N}$, so that distances are normalized in the same way as before, i.e., one vertex per unit area. Following the standard ``stub'' method \cite{Newman}, we attach $2M=400$ stubs (half-lines) randomly to the $N$ vertices. Then stubs are connected randomly in pairs, excluding configurations with disconnected components, as well as self-loops (two mutually connected stubs belonging to the same vertex). 500 random realizations are generated as our initial random-graph ensemble.

\subsection{Random Florida Grid I. $G(N,M,L,a)$}

\begin{figure}
\begin{center}
\hspace*{-2.5cm}\includegraphics[height=9cm, trim=0cm 0cm 0cm 0.1cm, clip=true]{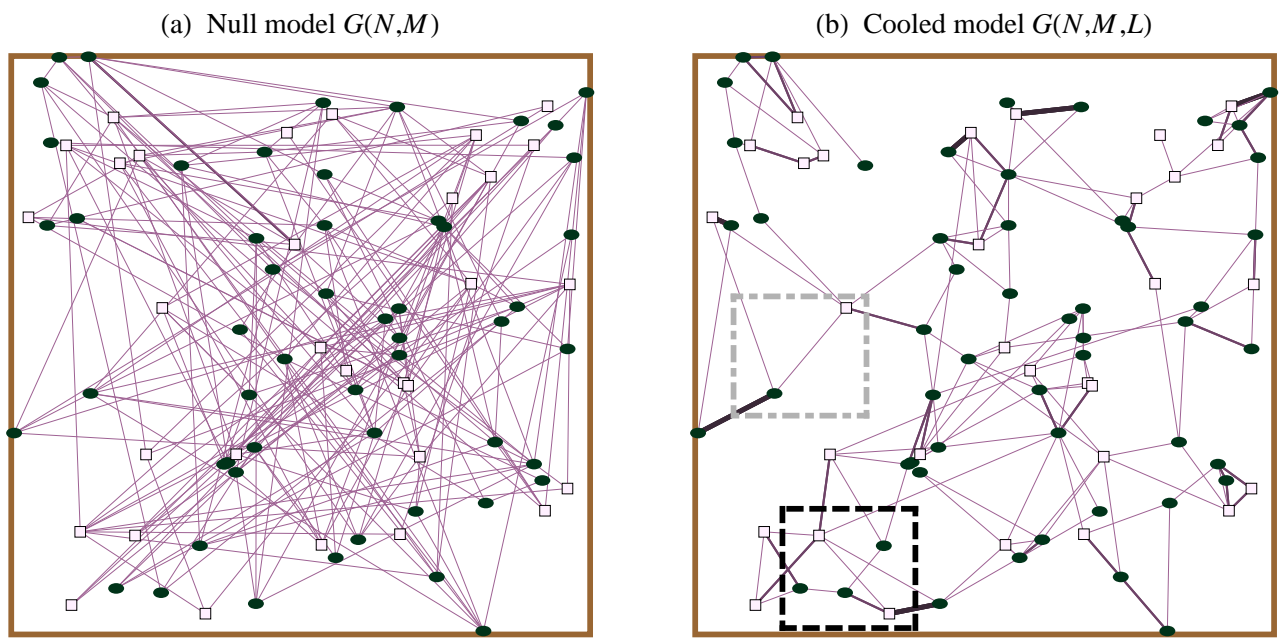}
\end{center}
\caption{\label{cooling} The Monte Carlo (MC) cooling procedure illustrated for $a=1$. \mbox{(a) The null model} $G(N,M)$, in which $N_g=31$ generators and $N_l=53$ loads are distributed randomly and then randomly connected with $M=200$ power lines. \mbox{(b) The model $G(N,M,L)$} obtained from (a) by our MC cooling procedure with temperature $T=0.545$. The dashed boxes are referenced in Fig. \ref{final}. }

\end{figure}

Notice that there are many long-distance lines in $G(N,M)$ (Fig. \ref{cooling}(a)); this configuration is unrealistic due to obvious economic considerations. In order to fix this problem, a modified model was first proposed in Ref. \cite{Rikvold}, in which a Monte Carlo (MC) ``cooling'' procedure was employed. See Fig. \ref{cooling}(b) for illustration. We name this model $G(N,M,L)$ because the total length of transmission lines $L \equiv \sum_{ij}{L(ij)}$ plays the role of the system energy (``Hamiltonian'') in the MC cooling procedure. Here $L(ij)$ denotes the total length of all the direct transmission lines between stations $i$ and $j$. For two randomly chosen lines $ij$ and $kl$, we consider the rewired pair $il$ and $kj$, which corresponds to the change in total line length $\Delta L$. The new configuration is accepted with Metropolis probability, $P(ij,kl \rightarrow il,kj)=\rm{Min}[1, \rm{exp}(-\Delta \emph{L}/\emph{T})]$, where $T$ is a fictitious ``temperature.'' Clearly, this update rule cools the network down to a configuration of lower, and thus more realistic, average transmission line length. For a properly chosen temperature ($T=0.545$), random graphs generated by this model display a line-length distribution with the average very close to that of FLG, viz. $\langle \ell \rangle \approx 1.09$ in our dimensionless units. See Ref. \cite{Rikvold} for a detailed description. 


%

\begin{figure}
\begin{center}
\includegraphics[width=10cm]{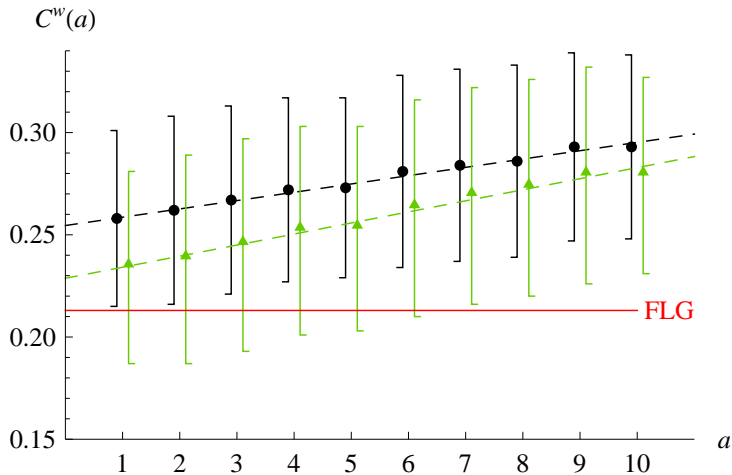}
\end{center}
\caption{\label{clustering}The weighted clustering coefficient $C^w$ as a function of aspect ratio $a$. $C^w(a)$ is  measured for $G(N,M,L,a)$ (circles) and $G(N,M,L,R_g,a)$ (triangles) with integer aspect ratios (plot points are slightly horizontally offset for readability).  The dashed lines are guides to the eye. The ``error bars'' are the empirical standard deviations $\sigma$, calculated over 500 independent realizations (of either model) for each aspect ratio.}
\end{figure}

Different administrative regions have different geographical shapes, which can be characterized most simply by their aspect ratios. For example, Egypt and the US state of New Mexico are like squares, while Turkey and the US state of California are more rectangular in shape. In particular, Chile is a country of very high aspect ratio. It is therefore of interest to embed models into areas with different aspect ratios, and investigate how geometric factors affect the basic metrics of power-grid networks. Here the model $G(N,M,L)$ is constructed in spaces of the same area as FLG (illustrated in Fig. \ref{FloridaGrid}(b)) but with 10 different aspect ratios, ranging from $a=1$ (square) to $a=10$ (long rectangle with a height-to-width ratio of 10), denoted as $G(N,M,L,a)$. For each aspect ratio, 500 independent realizations were generated by MC cooling. Our results, reported in Table \ref{Data}, are the averages and the empirical standard deviations $\sigma$ over the 500 configurations. Networks generated with this algorithm exhibit a large weighted clustering coefficient, i.e., $C^w(a) \approx 0.26$ for $G(N,M,L,a=1)$, which is much greater than $C^w \approx 0.052$ for the null model $G(N,M)$ with $a=1$, and slightly greater than (but close to) the real-world value $C^w_{\rm{FLG}} \approx 0.213$ (Fig. \ref{clustering}).  The reason for this phenomenon is that during the cooling process, long-distance lines are replaced by local shorter-distance connections, thus the cliquishness of local neighborhoods (measured by the clustering coefficient) significantly increases \cite{Spatial}. 

While capturing the clustering behavior of FLG, networks generated by $G(N,M,L,a)$ do not exhibit the desirable mixing pattern of generators and loads. For all aspect ratios, the mixing matrix for $G(N,M,L,a)$ is very close to what we have estimated for $[e_{ij}]_{\rm{rand}}$, because MC cooling does not distinguish between generators and loads, and thus always generates a random mixing pattern. In particular, $g-g$ connections are not suppressed (see the $e_{gg}$ column in Table \ref{Data}) and are also on average longer than those in FLG (see the $\langle \ell_{gg} \rangle$ column). Furthermore, compared to FLG, the number of extra parallel transmission lines in $G(N,M,L,a)$, e.g., $M_{ex} \approx 50$ for $a=1$, is quite low, and the total resistance $R$ is in general several standard deviations higher. Illustrating the competition between edge multiplicity and clustering, the relatively large number of multiple edges in FLG coincides with a lower weighted clustering coefficient. These comparisons demonstrate that there is nontrivial architectural structure in FLG, which cannot be fully captured by the model $G(N,M,L,a)$ optimized to match the total design cost of power lines. 

\subsection{Random Florida Grid II. $G(N,M,L,R_g,a)$}

Although the simple model $G(N,M,L,a)$ does not reproduce the structural organization of FLG in detail, it is still useful as a starting point in the search for optimization principles behind the architecture of FLG. In this section, we propose a more refined model, $G(N,M,L,R_g,a)$, which is again embedded in rectangular regions of area $N$ and aspect ratio $a$. According to the discussion in previous sections, $g-g$ connections should be suppressed while $g-l$ connections should be favored, without affecting $l-l$ connections. Moreover, a certain fraction of multiple edges in power grids can increase the robustness (or reliability) of vertex connectivity against random line failures. 

\begin{figure}
\begin{center}
\includegraphics[width=9cm,trim=0cm 0cm 0cm 1cm,clip=true]{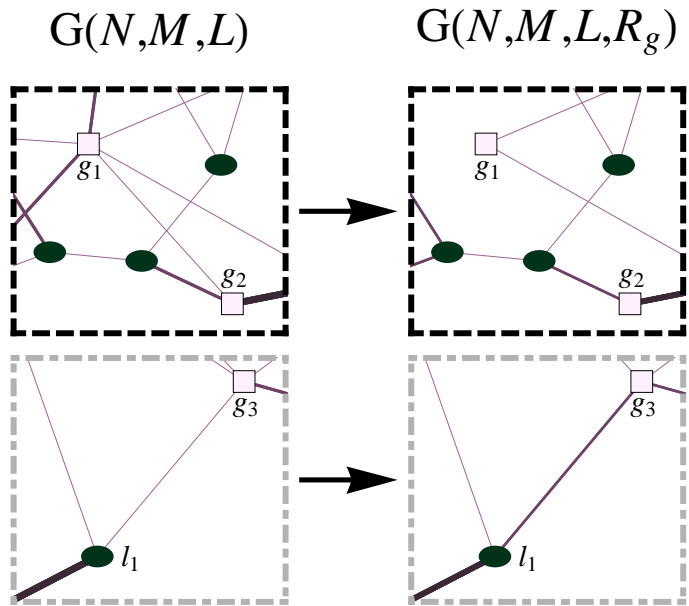}
\end{center}
\caption{\label{final} From  $G(N,M,L)$ to $G(N,M,L,R_g)$. Here we demonstrate the differences between $G(N,M,L)$ and $G(N,M,L,R_g)$, by comparing local neighborhoods of realizations generated by the two models. The neighborhood regions selected are those corresponding to the boxed regions in Fig. \ref{cooling}. In creating the new model, we remove long $g-g$ transmission lines (e.g., line $g_1-g_2$ in the dark dashed box) and replace them by adding multiples to $g-l$ edges of similar length to the ones removed (e.g., the multiple $g_3-l_1$ line added in the light dashed box).}
\end{figure}

Starting from $G(N,M,L,a)$, we performed the following heuristic optimization algorithm (see Figs. \ref{cooling} and \ref{final} for illustrations). Remove the longest $g-g$ transmission line, select the $g-l$ edge of the length closest to that of the $g-g$ line just removed, add an extra transmission line between the selected $g-l$ vertex pair, and repeat this replacement procedure until the total edge resistance closely matches that of FLG, as shown in the column of $R$ values in Table \ref{Data}. This choice of $g-l$ vertex pair (of the length closest to that of the $g-g$ line removed), leaves the length distribution of power lines (and thus their total length $L$) almost unaffected, while reducing the average resistance of $g-l$ pairs $\langle R_ {gl} \rangle$ (and thus the total resistance $R$). By replacing transmission lines in this way, a significant fraction of $g-g$ lines are removed, while the number of parallel $g-l$ lines increases, so both $R_{gg}$ and $R_{gl}$ (and thus the total edge resistance $R$) decreases towards the values in FLG, as seen in Table \ref{Data}.

Like the previous model $G(N,M,L,a)$, the current model $G(N,M,L,R_g,a)$ exhibits a high degree of clustering, but with values closer to FLG, as shown in Fig. \ref{clustering}. This is due to the increased number of multiple edges (recall the competition between clustering and edge multiplicity). We stress that in creating $G(N,M,L,R_g,a)$, only the total edge resistance $R$ is optimized to match the value in FLG. Though there is no artificial matching for other network metrics, it is interesting to observe that $G(N,M,L,R_g,a)$ exhibits satisfying results, e.g., the mixing matrix element $e_{gg}$, the average length of $g-g$ edges $\langle \ell_{gg} \rangle$, and the number of extra lines $M_{\rm{ex}}$. Such agreement of our model with FLG suggests a possible design principle giving rise to the architecture of FLG: the construction cost of transmission lines (measured by $L$) and the total edge resistance are both minimized to a certain extent, while at the same time keeping a significant level of clustering so that the grid connectivity is robust against random failures of stations and power lines.

For some network metrics, e.g., $e_{gg}$ and $R$, FLG can be captured by our model $G(N,M,L,R_g,a)$ with the $a$ value lying between $8$ and $10$, which is consistent with the shape of Florida as seen from the map in Fig. \ref{FloridaGrid}. However, for some other metrics like $\langle \ell_{gg}\rangle$ and $C^w$, the model $G(N,M,L,R_g,a)$ with $a$ near $1$ (square shape) shows a better agreement. This implies that aspect ratio is only a rough measure of the geometric features of power-grid networks. The reason is that power grids consist of stations and transmission lines, which are not two-dimensional objects from the network perspective, although they are spatially embedded in a plane. This motivates us to introduce a more precise geometric measure, the box-counting fractal dimension $d_B$, to characterize the spatial extension of power grids. 


\begin{center}

 \begin{table*}[ht]
{\small
\caption{\label{Data} Network properties of FLG, $G(N,M,L,a)$ and $G(N,M,L,R_g,a)$. $\langle\ell_{gg}\rangle$ is the average geographic length of $g-g$ transmission lines; $e_{gg}$ is the fraction of power lines connecting generators to generators directly; $C^w$ is the (edge conductance) weighted clustering coefficient; $R$ is the total edge resistance; $\langle R_{gg} \rangle$ is the average $g-g$ edge resistance; $\langle R_{gl} \rangle$ is the average $g-l$ edge resistance; $M_{\rm{ex}}$ is the number of parallel transmission lines forming multiple edges; $d_B$ is the box-counting fractal dimension of power grid networks defined in Section IV. The reported values are averages and the empirical standard deviations $\sigma$, calculated over 500 independent realizations for each aspect ratio.\vspace{0.5em}}
\hfill{}
 \bgroup
 \def\arraystretch{1.2}
\setlength{\tabcolsep}{3pt}
 \makebox[\linewidth][c]{ 
    \begin{tabular}{cccccccccc  }

\hline      
\multicolumn{1}{c}{Power Grids} &  \multicolumn{1}{c}{$\langle\ell_{gg}\rangle$} & \multicolumn{1}{c}{$e_{gg}$} & \multicolumn{1}{c}{$C^w$} & \multicolumn{1}{c}{$R$} & \multicolumn{1}{c}{$\langle R_{gg} \rangle$} & \multicolumn{1}{c}{$\langle R_{gl} \rangle$}  & \multicolumn{1}{c}{$M_{\rm{ex}}$} & \multicolumn{1}{c}{$d_B$} \\   \hline  \\[-0.5em]

\multicolumn{1}{c}{FLG}  & \multicolumn{1}{l}{$\approx 0.71$} & \multicolumn{1}{l}{$=0.085$} & \multicolumn{1}{l}{$\approx 0.21$} & \multicolumn{1}{l}{$\approx 139.78$} &  \multicolumn{1}{l}{$\approx 0.57$} & \multicolumn{1}{l}{$\approx 0.96$} & \multicolumn{1}{l}{$=63$}  & \multicolumn{1}{l}{$\approx 1.37$} \\ 
\\[-0.5em]
\multicolumn{1}{c}{$G(N,M,L,a)$} \\ 
$a=1$ &  $1.16 \pm 0.18$ & $0.13 \pm 0.03$ & $0.26 \pm 0.04$ & $173.01 \pm 13.61$ & $1.15 \pm 0.19$ & $1.15 \pm 0.09$ & $49.61 \pm 5.76$ & $1.55 \pm 0.02$ \\
$a=2$ &  $1.14 \pm 0.17$ & $0.13 \pm 0.03$ & $0.26 \pm 0.05$ & $173.09 \pm 13.42$ & $1.14 \pm 0.18$ & $1.15 \pm 0.10$ & $49.74 \pm 5.52$ & $1.56 \pm 0.02$ \\
$a=3$ &  $1.14 \pm 0.18$ & $0.13 \pm 0.03$ & $0.27 \pm 0.05$ & $169.49 \pm 12.70$ & $1.13 \pm 0.19$ & $1.13 \pm 0.09$ & $50.80 \pm 5.33$ & $1.53 \pm 0.02$ \\
$a=4$ &  $1.13 \pm 0.17$ & $0.13 \pm 0.03$ & $0.27 \pm 0.05$ & $168.49 \pm 13.51$ & $1.12 \pm 0.18$ & $1.13 \pm 0.09$ & $51.60 \pm 5.61$ & $1.51 \pm 0.02$ \\
$a=5$ &  $1.14 \pm 0.16$ & $0.13 \pm 0.03$ & $0.27 \pm 0.04$ & $167.26 \pm 11.79$ & $1.13 \pm 0.18$ & $1.13 \pm 0.09$ & $52.00 \pm 5.25$ & $1.47 \pm 0.01$ \\
$a=6$ &  $1.13 \pm 0.17$ & $0.13 \pm 0.03$ & $0.28 \pm 0.05$ & $163.96 \pm 13.33$ & $1.11 \pm 0.19$ & $1.12 \pm 0.09$ & $53.15 \pm 5.63$ & $1.45 \pm 0.02$ \\
$a=7$ &  $1.12 \pm 0.17$ & $0.13 \pm 0.03$ & $0.28 \pm 0.05$ & $162.90 \pm 12.79$ & $1.11 \pm 0.18$ & $1.11 \pm 0.09$ & $53.86 \pm 5.23$ & $1.42 \pm 0.02$ \\
$a=8$ &  $1.13 \pm 0.18$ & $0.13 \pm 0.03$ & $0.29 \pm 0.05$ & $160.93 \pm 12.37$ & $1.11 \pm 0.19$ & $1.10 \pm 0.09$ & $54.41 \pm 5.56$ & $1.40 \pm 0.02$ \\
$a=9$ &  $1.11 \pm 0.17$ & $0.13 \pm 0.03$ & $0.29 \pm 0.05$ & $159.31 \pm 12.67$ & $1.10 \pm 0.18$ & $1.10 \pm 0.09$ & $55.33 \pm 5.55$ & $1.38 \pm 0.02$ \\
$a=10$ &  $1.11 \pm 0.17$ & $0.13 \pm 0.03$ & $0.29 \pm 0.05$ & $156.94 \pm 12.76$ & $1.08 \pm 0.18$ & $1.09 \pm 0.09$ & $56.23 \pm 5.85$ & $1.37 \pm 0.02$ \\

\\[-0.5em]
\multicolumn{1}{c}{$G(N,M,L,R_{g},a)$} \\ 
$a=1$ &  $0.72 \pm 0.38$ & $0.04 \pm 0.04$ & $0.23 \pm 0.05$ & $145.78 \pm \hspace{0.4em}8.36$ & $0.66 \pm 0.38$ & $1.03 \pm 0.08$ & $63.83 \pm 4.39$ & $1.52 \pm 0.02$ \\
$a=2$ &  $0.73 \pm 0.45$ & $0.04 \pm 0.05$ & $0.24 \pm 0.05$ & $146.08 \pm \hspace{0.4em}8.83$ & $0.66 \pm 0.45$ & $1.03 \pm 0.09$ & $63.79 \pm 4.16$ & $1.54 \pm 0.02$ \\
$a=3$ &  $0.74 \pm 0.39$ & $0.05 \pm 0.05$ & $0.24 \pm 0.05$ & $143.83 \pm \hspace{0.4em}7.41$ & $0.66 \pm 0.39$ & $1.02 \pm 0.08$ & $63.80 \pm 4.30$ & $1.51 \pm 0.02$ \\
$a=4$ &  $0.79 \pm 0.37$ & $0.05 \pm 0.05$ & $0.25 \pm 0.05$ & $144.47 \pm \hspace{0.4em}8.36$ & $0.71 \pm 0.39$ & $1.03 \pm 0.08$ & $63.82 \pm 4.32$ & $1.49 \pm 0.02$ \\
$a=5$ &  $0.80 \pm 0.37$ & $0.06 \pm 0.05$ & $0.25 \pm 0.05$ & $142.88 \pm \hspace{0.4em}6.58$ & $0.70 \pm 0.39$ & $1.02 \pm 0.08$ & $64.18 \pm 4.39$ & $1.46 \pm 0.01$ \\
$a=6$ &  $0.82 \pm 0.35$ & $0.07 \pm 0.05$ & $0.26 \pm 0.05$ & $142.11 \pm \hspace{0.4em}5.90$ & $0.72 \pm 0.36$ & $1.02 \pm 0.08$ & $63.85 \pm 4.14$ & $1.43 \pm 0.01$ \\
$a=7$ &  $0.82 \pm 0.33$ & $0.07 \pm 0.05$ & $0.27 \pm 0.05$ & $141.39 \pm \hspace{0.4em}5.75$ & $0.72 \pm 0.34$ & $1.02 \pm 0.07$ & $64.21 \pm 3.84$ & $1.41 \pm 0.01$ \\
$a=8$ &  $0.84 \pm 0.35$ & $0.08 \pm 0.05$ & $0.27 \pm 0.05$ & $140.89 \pm \hspace{0.4em}5.21$ & $0.73 \pm 0.36$ & $1.01 \pm 0.08$ & $63.71 \pm 4.11$ & $1.39 \pm 0.01$ \\
$a=9$ &  $0.87 \pm 0.33$ & $0.08 \pm 0.05$ & $0.28 \pm 0.05$ & $140.64 \pm \hspace{0.4em}5.35$ & $0.77 \pm 0.35$ & $1.01 \pm 0.07$ & $64.17 \pm 4.09$ & $1.37 \pm 0.01$ \\
$a=10$ &  $0.87 \pm 0.31$ & $0.08 \pm 0.05$ & $0.28 \pm 0.05$ & $140.00 \pm \hspace{0.4em}5.00$ & $0.77 \pm 0.33$ & $1.01 \pm 0.08$ & $64.14 \pm 3.93$ & $1.36 \pm 0.01$ \\

\hline

\end{tabular}}
}
 \egroup
\hfill{}

\end{table*}

\end{center}

\section{Box-counting dimensions and aspect ratios}

\begin{figure}

\begin{center}
\hspace*{-1.5cm}\includegraphics[height=10cm, trim=0cm 0cm 0cm 0cm, clip=true]{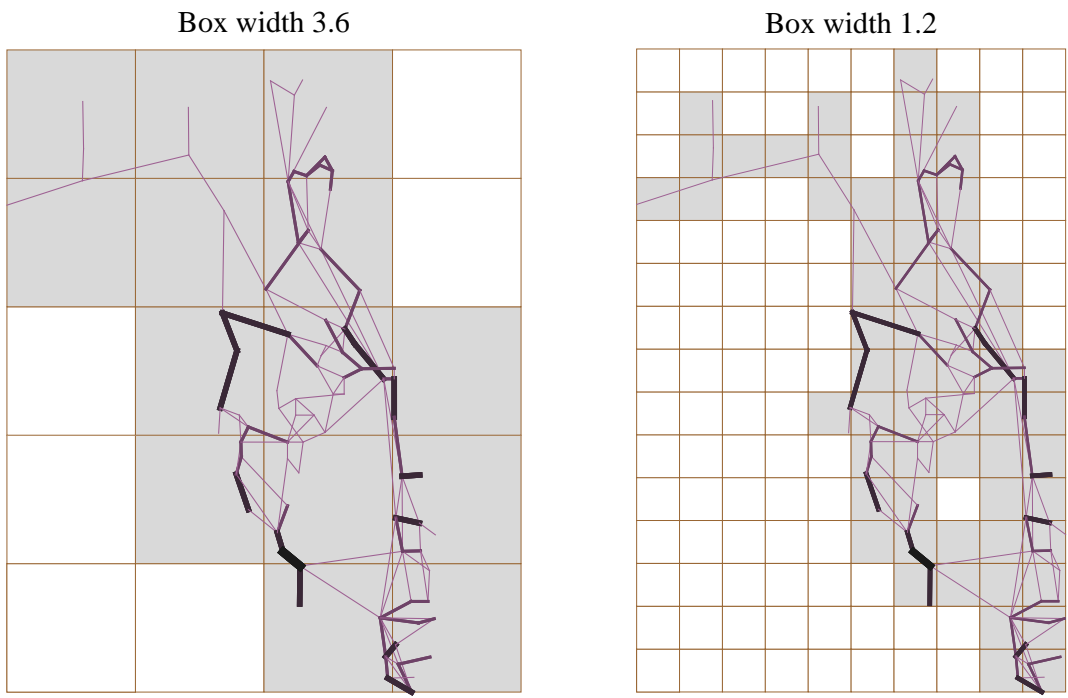}
\end{center}

\caption{\label{box} Box-counting illustrations for FLG. The shaded boxes intersect with the network object. The box widths are given in our dimensionless units. For a sense of length scale, recall that the average geographic length of transmission lines in FLG is $\langle\ell\rangle \approx 1.09$.}
\end{figure}

We now measure the box-counting dimension $d_B$ \cite{Feldman} of the models under consideration. The method partitions a given network using a grid of square boxes of side $r$, and counts the number of boxes $N(r)$ containing (part of) at least one edge (see illustrations in Fig. \ref{box}). For fractal network structures, $N(r)$ scales with the inverse magnification factor $1/r$ as a power law \cite{fractal}, thus $d_B$ is given by the equation 
\noindent
\begin{equation}
\mathrm{log} N(r) = c + d_B \,\mathrm{log}(1/r) .
\end{equation}
Because our graphs are not perfect fractals, we define $d_B$ to be the slope of the best-fit line of the points $\left\{\mathrm{log}N(r),\mathrm{log}(1/r)\right\}$. 

\begin{figure}
\begin{center}
\includegraphics[width=10cm,trim=0.1cm 0cm 0cm 0cm, clip=true]{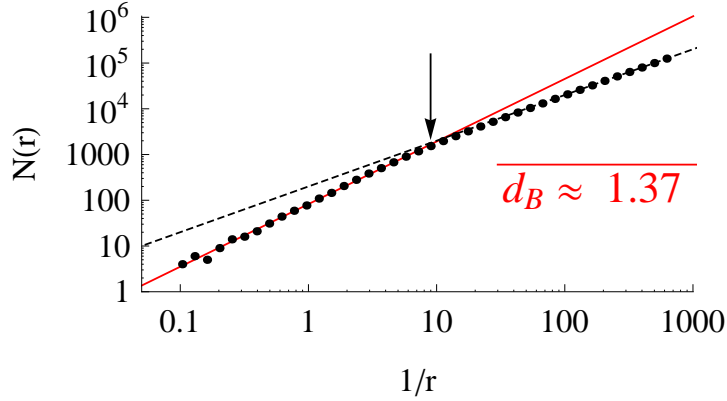}\hbox{\hspace{1.2cm}}
\end{center}
\caption{\label{fig:loglog} The box-counting log-log plot for FLG. The dashed line has slope $1$, which is the asymptotic limit of $d_B$ for all finite networks. The solid line corresponds to the reported value for FLG, $d_B \approx 1.37$, calculated as the best fit slope ($R^2 = 0.997$) in the interval between the point with the largest $r$ and that marked with an arrow. The endpoint corresponds to about $1/10$ of the average geographic length of power lines.}
\end{figure}

\begin{figure}
\begin{center}
\includegraphics[width=10cm]{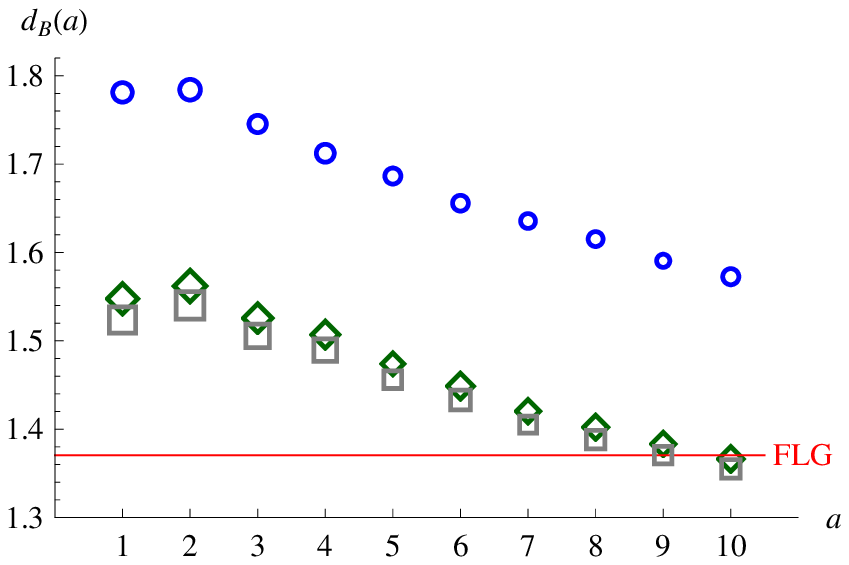}
\end{center}
\caption{\label{fractal} Box-counting dimension $d_B$ as a function of aspect ratio $a$.  $d_B(a)$ is measured for  $G(N,M,a)$ (circles), $G(N,M,L,a)$ (diamonds), and $G(N,M,L,R_g,a)$ (squares) at integer aspect ratios. The vertical extension of each data point indicates twice the empirical standard deviation $\sigma$, measured over 500 independent realizations (for each aspect ratio). For comparison, $d_B$ for the Florida Grid is plotted as the line labeled ``FLG".}
\end{figure}

Since power grids are finite collections of (one-dimensional) lines, the ``true" asymptotic value of $d_B$  is 1 when we consider boxes of vanishing size (the dashed line in Fig. \ref{fig:loglog}). However, when restricted to a finite range of box sizes, $d_B$ can be viewed as a measure of how effectively a set of edges fill their embedding space \cite{Batty}.  In Fig. \ref{fig:loglog}, we report that for FLG $d_B \approx 1.37$ over two decades of $r$ between $0.1$ and $10$.  This value is in the expected range: $d_B\ge1$ because the graph representing the grid is made up of lines, and $d_B< 2$ because the lines do not fully cover the embedding space. In Fig. \ref{fractal}, we compare the values of $d_B(a)$ measured for the null model $G(N,M,a)$, and the two optimized models, $G(N,M,L,a)$ and $G(N,M,L,R_g,a)$, each with 10 different aspect ratios. We see that both optimized models can take on the $d_B$ value for FLG when $a$ is near $9$, a value which is roughly consistent with the shape of FLG. In the aspect ratio range between $a=2$ and $a=10$, $d_B(a)$ is generally a non-increasing function of $a$. Also, $d_B(a)$ for $G(N,M,a)$ is significantly higher than that of both $G(N,M,L,a)$ and $G(N,M,L,R_g,a)$ (see the $d_B$ column in Table \ref{Data}), because in the uncooled model there are many long edges, which intersect many boxes even for relatively small $r$. Moreover, $d_B(a)$ for $G(N,M,L,R_g,a)$ is consistently lower (and thus closer to the real value) than that for $G(N,M,L,a)$ (see Fig. 7), since the power-grid networks generated by $G(N,M,L,R_g,a)$ have relatively fewer edges (due to the higher average edge multiplicities) and thus intersect fewer boxes.

Here we mention that there are actually two types of box-counting fractal dimensions that are relevant to spatially embedded networks such as FLG. What we have measured and presented here is inherently \emph{geometric}, revealing the spatial extension of the network in the (continuous) two-dimensional space. The other metric is \emph{topological} \cite{Song2007, Galvao2010}, which is reminiscent of the Real Space Renormalization Group (RSRG) approach in statistical physics, viz. grouping  vertices in the network according to their graphical distance (e.g., nearest-neighbors, second nearest-neighbors, etc.). Because this perspective on fractality measurement is topological rather than geographical, its analysis is heavily affected by the discrete nature of networks, and is also insensitive to the details of spatial configuration (embedding). It is quite difficult to perform reliable measurements of topological fractal dimension on small networks because the intrinsic discreteness of the graph structure prevents the accumulation of a sufficient number of data points \cite{Song2007a}. Therefore,  given that the FLG network is of relatively small size, viz. $84$ vertices, and that we here primarily focus on its geographical properties,  we only present results of the geometric box-counting analysis in this paper. However, in future studies it will be of interest to consider topological fractality in the context of network growth \cite{Makse1995}, and in particular the RSRG method of coarse-graining vertices of power grids into clusters of ``super-generators'' and ``super-loads" \cite{Hamad}.


\section{Conclusions}

Our measurements and modeling of power grids as spatial networks suggest that FLG has been organized such that the construction cost of transmission lines and the total edge resistance are both minimized to a certain extent, while at the same time keeping a significant level of clustering so that the grid connectivity is robust against random failures of stations as well as power lines. Also, compared to the random mixing of generators and loads, in FLG a significant fraction of $g-g$ connections are suppressed and replaced by $g-l$ connections, which allows more loads to get power supply directly from generators. Results indicate that our modeling reproduces a large number of the network metrics and thus uncovers the organizational principles behind the architecture of FLG. It would be of interest to extend this investigation and modeling to more real-world power-grid networks, which have been constructed in geographical regions with different geometric shapes. Particularly, from our analysis of a parametrized mixing matrix of generators and loads, we observed that the fraction $\alpha$ of $g-g$ power lines replaced by $g-l$ lines coincides with the density of generators $\rho_g$ in FLG, viz. $\alpha_{\rm{FLG}} \approx 0.375$ is close to $\rho_g \approx 0.369$. It is worth further investigating whether this exemplifies a universal pattern in power-grid networks, or is unique to FLG . We leave this for future research, as more data becomes accessible to us. 

Our optimization of power grids employs a rather global perspective, viz. vertices (generators and loads) and transmission lines are first placed randomly in space, and then individual transmission lines are rewired according to certain global optimization principles, e.g., reducing the total length of transmission lines or matching the total pairwise edge resistance. Recently, some authors have proposed that, by modeling spatial networks via global optimization principles, one overlooks the usually limited time horizon of planners and the local self-organization underlying the network's formation \cite{Louf}. Thus another direction to extend our analysis and modeling is to study the growth of spatial networks \cite{Barrat2005, Newman2006} under certain local rules, e.g., the emergence of hierarchy in cost-driven growth of spatial networks \cite{Louf}. 

\section*{Acknowledgments}
We are grateful to Gregory Brown for helpful discussions, and Gregory Brown and Svetlana V. Poroseva for critical reading of the manuscript. This work was supported in part by U.S. National Science Foundation Grant No. DMR-1104829.

\end{document}